\documentclass[aps,prd,preprint,preprintnumbers]{revtex4}
\usepackage{graphicx}

\newcommand{\cm}{{cm}}
\newcommand{\lab}{{lab}}

\begin{document}

\preprint{MPP-2007-15}

\title{Sensitivity  of low energy neutrino experiments to physics beyond 
the Standard Model}

\author{J. Barranco}
\email{jbarranc@fis.cinvestav.mx}
\author{O. G. Miranda}
\email{Omar.Miranda@fis.cinvestav.mx}

\affiliation{Departamento de F\'{\i}sica, Centro de Investigaci{\'o}n y de
  Estudios Avanzados del IPN, Apartado Postal 14-740 07000
  M\'exico, D F, M\'exico}

\author{T. I. Rashba}
\email{timur@mppmu.mpg.de}

\affiliation{Max-Planck-Institut f\"ur
Physik (Werner-Heisenberg-Institut), F\"ohringer Ring 6, 80805
M\"unchen, Germany}

\affiliation{Institute of Terrestrial Magnetism,
  Ionosphere and Radio Wave Propagation of the Russian Academy of Sciences,
  142190, Troitsk, Moscow region, Russia}

\date{\today}
\begin{abstract}
  We study the sensitivity of future low energy neutrino experiments to extra
  neutral gauge bosons, leptoquarks and $R$-parity breaking interactions. We
  focus on future proposals to measure coherent neutrino-nuclei scattering and
  neutrino-electron elastic scattering.  We introduce a new comparative
  analysis between these experiments and show that in different types of new
  physics it is possible to obtain competitive bounds to those of present and
  future collider experiments. For the cases of leptoquarks and $R$-parity
  breaking interactions we found that the expected sensitivity for most of the
  future low energy experimental setups is better than the current
  constraints.
\end{abstract}

\pacs{13.15.+g,12.90.+b,23.40.Bw}

\maketitle

\section{Introduction}
The Standard Model (SM) is one of the most successful models in physics
and it is in very good agreement with almost every measurement in high
energy physics\cite{Yao:2006px}. Despite this fact, there are many
motivations to believe that the SM is not the last step in the
description of the physics of elementary particles. 

There are many theoretical motivations to believe that there is
physics beyond the standard model, and recently the neutrino
oscillation experiments have also given an experimental input on these
thoughts. 
Among the most popular extensions of the SM we find grand unified
theories (GUT), supersymmetry (SUSY), and extra dimensions. None of these
theories have been observed in the laboratory, but there are extensive
searches for signatures of them in collider physics.
The main aim of this paper is to analyze the potential of low energy
neutrino experiments either to confirm the presence of new physics if
it would be discovered by the Large Hadron Collider (LHC), or put
stronger or complementary constraints on their parameters.

We center our attention in signatures that could appear in two
different reactions: coherent neutrino-nuclei scattering and
neutrino-electron elastic scattering.  As concrete examples of
coherent neutrino-nuclei scattering we will consider the TEXONO
proposal~\cite{Wong:2005vg}, a stopped pion source (SPS) with a noble
gas detector~\cite{Scholberg:2005qs} and the recently discussed
proposal of low energy beta beams~\cite{Bueno:2006yq}.  For the
neutrino electron scattering case, we concentrate in the Double Chooz
proposal~\cite{Ardellier:2006mn}.

For some of these experimental proposals there have already been
discussions about their perspectives for constraining non-standard
neutrino interactions~\cite{Barranco:2005yy,Scholberg:2005qs} or a
non-zero neutrino magnetic
moment~\cite{Scholberg:2005qs,McLaughlin:2003yg,deGouvea:2006cb,Wong:2005pa}.
In this work we introduce a new comparative analysis between different low
energy experiments, focusing on three different types of new physics
phenomenology, namely extra neutral gauge bosons, leptoquarks and $R$-parity
breaking Supersymmetry. 
As far as we know, this is the first time that the sensitivity of low
energy neutrino proposals to leptoquarks is studied. On the other
hand, extra neutral gauge bosons sensitivity had been studied only for
the TEXONO and neutrino electron scattering
proposals~\cite{Barranco:2005yy,deGouvea:2006cb,Moretti:1998py}.  For
the case of $R$-parity breaking Supersymmetry the existing studies
have tested either long-baseline neutrino
experiments~\cite{Adhikari:2006uj} that introduce an extra dependence
on $\theta_{13}$, or new physics effects in the source due to charged
currents~\cite{Agarwalla:2006ht}, while here we will focus on neutral
currents effects, visible in the detector, specifically in a short
baseline detector based on coherent neutrino nuclei scattering.
Moreover, the study of different future proposals at one time gives to
the reader an extra usefulness of telling which future experiments
will give better chances in the different types of new physics under
study.
We will see that, despite the fact that we are dealing with very low
energy experiments there are good chances to obtain a very good
sensitivity to these types of new physics and either to compete or to
give complementary constraints to those that could be obtained from
collider experiments.

The structure of the article is the following: In section II we
describe the experimental proposals that we study. In section III we
introduce the different types of new physics under consideration and
the expected sensitivity in the different experimental setups. Finally
in section IV we present our conclusions. 

\section{Experimental proposals}
Before introducing the phenomenology to study new physics signatures
we would like to discuss the low energy neutrino experimental
proposals. In particular, we will discuss the case of future
experiments aiming to measure the coherent neutrino scattering off
nuclei as well as the case of low energy neutrino-electron-scattering
experiments. For the first reaction we study three different recent
proposals while for neutrino-electron scattering we concentrate in the
Double Chooz case.

\subsection{TEXONO}

TEXONO collaboration has recently started a research program towards
the measurement of neutrino-nuclei coherent scattering by using
reactor neutrinos and an ``ultralow-energy'' germanium detector
(ULEGe)~\cite{Wong:2005vg}.

The proposed detector would consist of 1~kg of an ultralow-energy
germanium detector with a threshold as low as 100~eV and a background
level below 1~keV in the range of 1 count-per-day that implies a
signal to noise ratio bigger than 22. Although an estimate for the
systematic uncertainties is not available, we can consider that they
will be dominated by the reactor power, its fuel composition, and the
anti-neutrino spectrum. We assume that these uncertainties will give
an approximate error of 2\%~\cite{Huber:2004xh}.

Besides the 100~eV threshold, we will also consider the more
conservative case of a 400~eV threshold. The typical time scale of
data taking is assumed to be from one to several years.

The electron anti-neutrino flux is coming from the Kuo-Sheng Nuclear
Power Station. The detector will be located at a distance of 28~m from
the reactor core. In our computation we will assume a typical reactor
neutrino flux of $10^{13}$~s$^{-1}$~cm$^{-2}$.  There are several
parameterizations that consider in detail the neutrino spectrum coming
from a reactor~\cite{Huber:2004xh,Schreckenbach:1985ep,Vogel:1989iv}.
In this work we will use the most recent
parameterization~\cite{Huber:2004xh} for the neutrino spectrum. Since
the proposed experiments are not running yet, we will assume that the
relative contribution of the fissile isotopes ($^{235}$U, $^{239}$Pu,
$^{238}$U, $^{241}$Pu) is given by the typical average values of the
reactor operating period~\cite{Kopeikin:1997ve} which is given by
$0.58:0.30:0.07:0.05$. We have checked numerically that the results
does not change significantly with other ratios. For energies below
$2$~MeV there are only theoretical calculations for the antineutrino
spectrum that we take from Ref.~\cite{Kopeikin:1997ve}.

Since we are not able to account for the detector efficiency and
resolution, we will estimate the total number of expected
events in a detector as
\begin{equation}
N_{\rm{events}}^{TEXONO}=t\,\phi_0\,\frac{M_{\rm{detector}}}{M}
\int\limits_{0}^{E_{max}}dE_\nu
\int\limits_{T_{th}}^{T_{max}(E_\nu)}dT\,
\lambda(E_\nu)\,\frac{d\sigma}{dT}(E_\nu,T)\,,
\label{Nevents}
\end{equation}
with $t$ the data taking time period, $\phi_0$ the total neutrino
flux, $M_{\rm{detector}}$ the total mass of the detector,
$\lambda(E_\nu)$ the normalized neutrino spectrum, $E_{max}$ the
maximum neutrino energy, $T_{th}$ the detector energy threshold. The
maximal recoil energy is $T_{max}(E_\nu)=2E_\nu^2/(M+2E_\nu)$. The
same expression relates the minimum required incoming neutrino energy with
the detector threshold $T_{th}$.  For instance, for the detector's
threshold 400~eV and $^{76}$Ge nucleus, the minimum required incoming
neutrino energy is about 3.8~MeV which is well satisfied for reactor
neutrinos.

\subsection{Stopped pion neutrino source}

A different proposal for detecting the coherent neutrino-nucleus
scattering considers the use of another source of neutrinos, a SPS,
such as the Spallation Neutron Source at Oak Ridge National
Laboratory.  Recently, this type of source was proposed to measure
coherent neutrino scattering off nuclei as well as non-standard
neutrino properties~\cite{Scholberg:2005qs}.

The total beam flux consists of the following well-known neutrino fluxes:
\begin{itemize}
\item the monoenergetic 29.9~MeV $\nu_\mu$'s produced from pion decay
  at rest, $\pi^+\to\mu^+ \nu_\mu$;
\item $\bar\nu_\mu$ and $\nu_e$  coming from muon decay, $\mu^+\to e^+\nu_e
  \bar\nu_\mu$, with a time delay about $2.2\mu$s, muon decay time scale.
\end{itemize}
The neutrino spectra are well known.  Here we will consider only the total
delayed flux ($\nu_e$~$+$~$\bar{\nu_\mu}$) as was done in
Ref.~\cite{Scholberg:2005qs}. We assume a total flux of
$\sim10^7\nu$~s$^{-1}$~cm$^{-2}$. Among different possible detector materials
such as Ar, Ge or Xe, we will concentrate on the noble gas detector,
$^{20}$Ne, of typical mass about $100$~kg with a data taking time scale from
one to several years and a threshold of $10$~keV.

\subsection{Low energy beta beams}

The usage of accelerated radioactive nuclei to produce a well known
flux of neutrinos -- beta beam -- was proposed
in~\cite{Zucchelli:2002sa}. It was shown soon afterwards that low
energy beta beams open new possibilities to study neutrino
properties~\cite{Volpe:2003fi} and, recently, a neutrino-nuclei coherent
scattering experiment using neutrinos from low energy beta beams was
discussed~\cite{Bueno:2006yq}.  On the other hand, tests for $R$
parity violating Supersymmetry have been discussed both by the direct
detection of $\tau$ leptons in a nearby
detector~\cite{Agarwalla:2006ht} as well as in long baseline beta-beam
experiments~\cite{Adhikari:2006uj}.

In particular we base our analysis on the beta-beam experiment discussed
in~\cite{Serreau:2004kx,Bueno:2006yq}.
We consider a storage ring of total length $L=1885$~m with a straight
sections of length $D=678$~m. In the stationary regime the mean number
of ions in the storage ring is $\gamma\tau g$, where $\tau=t_{1/2}/\ln
2$ is the lifetime of the parent nuclei, $g=2.7\times 10^{12}$ is the
number of injected ions per second and $\gamma=1/\sqrt{1-\beta^2}$ is
the time delay factor with $\beta$ the ion velocity in the laboratory
frame. As previous authors, we will consider a cylindrical detector of
radius $R=52$~cm and depth $h=40$~cm, aligned with one of the storage
ring's straight sections, and located at a distance $d=10$~m from
it. Integration over the decay path and over the volume of the
detector gives the total number of events per unit time
\begin{equation}
\label{dNevdt}
  N_{events}^{\beta-beam}=t\, g\,\tau\, n\,h\times
  \int_0^\infty dE_\nu\,\Phi_{tot}(E_\nu)\,\sigma(E_\nu)\,,
\end{equation}
where $t=1$~year is the data taking time, $n$ is the number of target nuclei
per unit volume, $\sigma(E_\nu)$ is the relevant neutrino-nucleus
cross-section. For definiteness we consider the case of a ton of Xe as a
target and a factor $\gamma=14$ for $^6$He ions as described in
Ref.~\cite{Bueno:2006yq}.  As for the threshold energy, we consider both the
realistic threshold of $15$~keV where background events are negligible as well
as the very optimistic $5$~keV threshold that, according to the same
reference, will give a bigger number of events if background could be
subtracted, though at present there is no technology capable of dealing with
such a background.  The total neutrino flux through detector is given by
\begin{equation}
\label{phitot}
\Phi_{tot}(E_\nu)=\int_0^D \frac{d\ell}{L}\,\int_0^h \frac{dz}{h}\,
\int_0^{\bar{\theta}(\ell,z)} \frac{\sin\theta d\theta}{2}\,
\Phi_{lab}(E_\nu,\theta)\,,
\end{equation}
where
\begin{equation}
\label{theta}
  \tan\bar{\theta}(\ell,z)=\frac{R}{d+\ell+z}\,.
\end{equation}
The boosted flux in the laboratory frame is
\begin{equation}
\label{bflux}
  \Phi_\lab(E_\nu,\theta) = \frac{\Phi_\cm(E_\nu\gamma[1-\beta\cos\theta])}
  {\gamma[1-\beta\cos\theta]}\,,
\end{equation}
where $E_{\nu}$ and $\Omega\equiv(\theta,\varphi)$ denote the energy
and solid angle of the emitted (anti-)neutrino in the laboratory
($lab$) frame and $\theta$ denotes the angle of emission with respect
to the beam axis.

The neutrino flux in the rest frame, $\Phi_\cm (E_\nu')$, is given by the
well-known formula~\cite{krane}:
\begin{equation}
\label{e:2}
  \Phi_\cm (E_\nu')=\frac{\ln 2}{m_e^5 ft_{1/2}}\,(E_\nu')^2\,E_e\,
  \sqrt{E_e^2-m_e^2}\, F(\pm Z,E_e)\,\Theta(E_e-m_e)\,.
\end{equation}
where $m_e$ the electron mass and $ft_{1/2}$ the {\it ft} value. The energy of
emitted lepton (electron or positron) is $E_e=Q-E_{\nu}'$, where $Q$ is the
$Q$ value of the reaction, and the Fermi function $F(\pm Z,E_e)$ accounts for
the Coulomb modification of the spectrum~\cite{feister}.

\subsection{Reactor experiments}
A different type of experiment that we will also consider in this article is 
based on low energy neutrino-electron scattering. This process has already
been considered as a possible place to search for an extra gauge
boson~\cite{Miranda:1997vs,Moretti:1998py}. The case of a reactor source to
constrain new physics has recently been discussed both for
present~\cite{Barranco:2005ps} and future proposals~\cite{deGouvea:2006cb}. In
this work we will concentrate on the perspectives for the Double Chooz
experiment~\cite{Ardellier:2006mn}.  As in~\cite{deGouvea:2006cb}, we assume
that the Double Chooz will collect $10^4$ neutrino-electron-scattering events
considering a 3GW reactor and a 26.5 ton detector with an electron visible
energy window $3$~MeV~$<T<5$~MeV. As in the case of the TEXONO proposal, we
will use the most recent parameterization~\cite{Huber:2004xh} for the neutrino
spectrum and the same fuel composition.

\subsection{Discussion on experiments}
We summarize the main characteristics of the detectors in
Table~\ref{detectors}. One can notice that in some cases it could be
possible to run the experiment for a period longer than one year, or
to upgrade the detector mass, obtaining a smaller statistical error
without being dominated by systematic uncertainties. This is the case,
for example, for a beta beam with a 15 keV threshold. On the other
hand the Stopped pion source seems to be suitable only for a one year
of data taking. Finally we also consider the very optimistic cases in
which experimentalists can reduce the uncertainties in a low threshold
regime (like a beta beam with a 5 keV threshold). In this case we
assume that the systematic uncertainties remain the same.

In the next sections we will take into account all these experimental
setups.  We will also show results for possible upgrades to these
experiments, i.e., we will consider that the experimental setup can
be running for a longer time (or that an upgrade in mass is possible).
Among the difficulties for the upgrade we must take special care of
the systematic error expectations.  In order to take a reasonable
compromise with future experimental capabilities, we will consider the
systematic errors quoted in Table~\ref{detectors}.  Since we are
dealing with experiments that are not running yet, we believe that
this approach will be helpful to take notice of what would be the
expected limits for each experiment.

\begin{table}[!t]  
    \begin{center}    
        \begin{tabular}{|c|c|c|c|}
            \hline     
 Experiment  & $M_0$ &Expected events/yr& systematic error estimate \\
\hline 
 Texono, $E_{th}=$400 eV    & 1 kg, Ge    & 3790  & 2 \%  \\ 
 Texono, $E_{th}=$100 eV    & 1 kg, Ge    & 25196 & 2 \%  \\ 
 Beta beam, $E_{th}=$15 keV & 1 ton, Xe    & 1390  & 2 \%  \\ 
 Beta beam, $E_{th}=$5 keV  & 1 ton, Xe    & 5309  & 2 \%  \\ 
 Stopped pion, $E_{th}=$10 keV  & 100kg, Ne    & 627   & 5 \%  \\  
 Double Chooz              & 26.5 tons, scintill. & 10000 & 1 \%  \\
 \hline
        \end{tabular}  
    \end{center}  
    \vskip -0.2cm  
    \caption{Expected events for different experimental setups}
    \label{detectors}
\end{table}

\section{Models and sensitivity}
Once the experimental setups have been discussed, we turn our attention to
different types of new physics that could be constrained in these future
proposals. We will consider three different scenarios that will be discussed
in detail in the following subsections.

\subsection{$Z'$ models}
In this section we introduce the description of the extra gauge bosons
to be considered. 
New massive gauge bosons are a common feature of
physics beyond the Standard Model.
Heavy neutral vector bosons $Z'$ are predicted in
string inspired extensions of the SM, in left-right symmetric models,
in models with dynamical symmetry breaking, in "little Higgs" models
and in certain classes of theories with extra dimensions. In many of
these models it is expected that $Z'$ mass can be around TeV scale.

The present experimental lower limits to the neutral gauge boson mass
come from the Tevatron and LEP experiments~\cite{Yao:2006px}.
Forthcoming measurements at LHC will provide sensitivity to the $Z'$
mass up to 5~TeV~\cite{Dittmar:2003ir,Rizzo:2006nw}.

The new $Z'$ boson affects the neutral current couplings of the SM, and its
contribution at low energies can be tested from atomic parity violation and by
electron-nucleon scattering experiments (see refs in~\cite{Yao:2006px}). Since
low energy experiments are not sensitive to the mixing angle between the SM
gauge boson and the extra gauge boson, and this angle is very well
constrained~\cite{Yao:2006px}, we will neglect it.

We consider first the particular case of an additional neutral gauge
boson $Z'$ that arises from a primordial $E_6$ gauge
symmetry~\cite{Gonzalez-Garcia:1990yq}.  These extension usually
involve an extra $U(1)$ hypercharge symmetry at low energies that may
be given as the mixture of those associated with the symmetries
$U(1)_\chi$ and $U(1)_\psi$.  We show the quantum numbers for the SM
particles in Table~\ref{tab:E6}.

The corresponding hyper-charge is then specified by
\begin{equation}
Y_\beta =Y_{\chi}\cos \beta +Y_{\psi}\sin \beta, 
\end{equation}
while the charge operator is given as $ Q=T^3+ Y $.  Any value of
$\beta$ is allowed, giving us a continuum spectrum of possible models
of the weak interaction.
At tree level it is possible to write an expression for the effective
4-fermion Lagrangian describing low-energy neutral current phenomena. We
neglect nonstandard radiative corrections because its contribution is of
order $(\alpha/\pi)(M_Z^2/M_{Z'}^2)$~\cite{Cvetic1987}.  
\begin{table}
    \caption{Quantum numbers for the light particles in the {\bf 27} of $E_6$.}
    \label{tab:E6}
        \begin{tabular}{cccc}
            \hline {\rule[-3mm]{0mm}{8mm}} & $T_3$ & $\sqrt{40}Y_\chi$ &
            $\sqrt{24}Y_\psi$   \\
            \hline
            $Q$   & $\pmatrix{1/2 \\ -1/2}$ & $-1$ & $1$ \\
            $u^c$ &       $0$               & $-1$ & $1$ \\
            $e^c$ &       $0$               & $-1$ & $1$ \\
            $d^c$ &       $0$               & $ 3$ & $1$ \\
            $l$   & $\pmatrix{1/2 \\ -1/2}$ & $ 3$ & $1$ \\
            \hline
        \end{tabular}  
\end{table}
Another class of $Z'$ models is coming from left-right symmetric
models that have the premise that the fundamental weak interaction
Lagrangian is invariant under parity symmetry at energies about 100
GeV. The gauge group of this type of models is given by
$SU(2))_L\otimes SU(2)_R\otimes U(1)_{B-L}$, which gives an additional
neutral gauge boson plus a charge gauge
boson~\cite{Erler:1999ub,Mohapatra-book}.  We will concentrate in this
work on the neutral currents.

In the following subsections we will introduce the modifications to
the coupling constants, and therefore to the cross section, due to
this type of new physics. With this information we will study the
different experimental proposals and their sensitivity to both $E_6$
and left-right symmetric neutral gauge bosons.

\subsubsection{Coherent neutrino-nuclei scattering coupling constants}

Before introducing this description it is useful to recall the general
description of the nonstandard neutrino-quark and neutrino-electron
interactions and then we will specify the interactions for commonly
used $Z'$ models.

Generically the neutrino-quark interaction at low energies (energies $\ll
M_Z$) can be described at the 4-fermion approximation by the effective
Lagrangian
\begin{equation}
\label{lagrangian}
{\cal L}_{\nu Hadron}^{NC}=-\frac{G_F}{\sqrt{2}}
\sum_{{q=u,d}}
\left[\bar\nu_e \gamma^\mu (1-\gamma^5)\nu_e \right]
\left(
f^{qL}\left[\bar q\gamma_\mu (1-\gamma^5) q\right] +
f^{qR}\left[\bar q\gamma_\mu (1+\gamma^5) q\right] 
\right),
\end{equation}
where 
\begin{eqnarray}
&&
f^{uL}=\rho_{\nu N}^{NC}
\left(
\frac12-\frac23 \hat\kappa_{\nu N} \hat s_Z^2
\right) + \lambda^{uL}+ \varepsilon^{uL},\nonumber\\
&&
f^{dL}=\rho_{\nu N}^{NC}
\left(
-\frac12+\frac13 \hat\kappa_{\nu N} \hat s_Z^2
\right) + \lambda^{dL}+ \varepsilon^{dL},\nonumber\\
&&
f^{uR}=\rho_{\nu N}^{NC}
\left(
-\frac23 \hat\kappa_{\nu N} \hat s_Z^2
\right) + \lambda^{uR}+ \varepsilon^{uR},
\label{couplings}\\
&&
f^{dR}=\rho_{\nu N}^{NC}
\left(
\frac13 \hat\kappa_{\nu N} \hat s_Z^2
\right) + \lambda^{dR}+ \varepsilon^{dR}\nonumber
\end{eqnarray}
Here $\hat s_Z^2=\sin^2\theta_W=0.23120$ -- the Weinberg weak mixing
angle taken in the $\bar{\mbox{MS}}$ model. The radiative
corrections~\cite{Yao:2006px} $\rho_{\nu N}^{NC}=1.0081$,
$\hat\kappa_{\nu N}=0.9978$, $\lambda^{uL}=-0.0031$,
$\lambda^{dL}=-0.0025,$ and
$\lambda^{dR}=2\lambda^{uR}=7.5\times10^{-5}$ are included into our
analysis.
In general, the parameters $\varepsilon^{qP}$ ($q=u,d$ and $P=L,R$)
describe a generic nonstandard neutrino interaction. For the specific
case of $E_6$ string inspired models this is translated into
\begin{eqnarray}
\label{couplingsnewz}
\varepsilon^{uL}&=&- 4 \gamma \sin^2\theta_W \rho_{\nu N}^{NC}
\left({c_\beta \over \sqrt{24}}-{s_\beta \over 3}\sqrt{5 \over 8} \right)
\left({3 c_\beta \over 2 \sqrt{24}}+{s_\beta \over 6}\sqrt{5 \over 8} \right)
\nonumber \\
\varepsilon^{dR}&=&-8 \gamma \sin^2\theta_W \rho_{\nu N}^{NC}
\left({3 c_\beta \over 2 \sqrt{24}}+
{s_\beta \over 6}\sqrt{5 \over 8} \right)^2 ,\nonumber \\
\varepsilon^{dL}&=&\varepsilon^{uL} = -\varepsilon^{uR}, 
\end{eqnarray}
where $c_\beta=\mbox{cos}\beta$, $s_\beta=\mbox{sin}\beta$ and
$\gamma=(M_Z/M_{Z'})^2$.  
Three main models have been extensively studied, namely: the $\chi$
model ($\mbox{cos}\beta=1$), the $\psi$ model ($\mbox{cos}\beta=0$)
and the $\eta$ model (cos$\beta=\sqrt{3/8}$). In previous
articles~\cite{Miranda:1997vs} it has been stressed that low energy
neutrino experiments are more sensitive to the $\chi$ model than to
other $E_6$ models. However, for comparison with the expected
sensitivity to $Z'$ mass in different models at LHC we will consider a
continuum spectrum of possible models over parameter $\beta$.

From the Lagrangian in Eq.~(\ref{lagrangian}) we can obtain the 
coherent neutrino-nucleus differential cross section which is given
by
\begin{eqnarray}
\frac{d\sigma}{dT}&=&\frac{G_F^2 M}{2\pi}\left\{
(G_V+G_A)^2+\left(G_V-G_A\right)^2\left(1-\frac{T}{E_\nu}\right)^2-
\left(G_V^2-G_A^2\right)\frac{MT}{E_\nu^2}
\right\}\,,\label{diff:cross:sect}
\end{eqnarray}
where $M$ is the mass of the nucleus, $T$ is the recoil nucleus
energy, which varies from 0 to $T_{max}=2E_\nu^2/(M+2E_\nu)$, $E_\nu$
is the incident neutrino energy and
\begin{eqnarray}
\label{GV}
G_V&=& 
\left[\left(g_V^p+2\varepsilon_{ee}^{uV}+\varepsilon_{ee}^{dV}\right)Z+
\left(g_V^n+\varepsilon_{ee}^{uV}+2\varepsilon_{ee}^{dV}\right)N\right]
F_{nucl}^V(Q^2)\,,\\
G_A&=& 
\left[\left(g_A^p+2\varepsilon_{ee}^{uA}+\varepsilon_{ee}^{dA}\right)
\left(Z_+-Z_-\right)+
\left(g_A^n+\varepsilon_{ee}^{uA}+2\varepsilon_{ee}^{dA}\right)
\left(N_+-N_-\right)\right]
F_{nucl}^A(Q^2)\,.
\label{GA}
\end{eqnarray}
$Z$ and $N$ represent the number of protons and neutrons in the
nucleus, while $Z_{\pm}$ ($N_\pm$) stands for the number of protons
(neutrons) with spin-up and spin-down respectively.
From Eq.~(\ref{GA}) it is possible to see that the axial couplings
will vanish for even-even nuclei considered below.

The vector and axial nuclear form factors, $F_{nucl}^V(Q^2)$ and
$F_{nucl}^A(Q^2)$, are usually assumed to be equal and of order of
unity in the limit of small energies, $Q^2\ll M^2$. 
In our computations, for the sake of completeness we take into account the
vector form factor given in Ref.~\cite{Engel:1991wq}. We have also made our
computations taking into account previous calculations of this form
factor~\cite{Ahlen:1987mn,Freese:1987wu,Sehgal:1986gn}, and we found that there
is no difference in our results for both of them, which give us confidence to
consider that this theoretical estimation will not have an impact on the
systematic errors.
The SM neutral current vector couplings of neutrinos with
protons, $g_V^p$, and with neutrons, $g_V^n$, are defined as
\begin{eqnarray}
&&g_V^p=\rho_{\nu N}^{NC}\left(
\frac12-2\hat\kappa_{\nu N}\hat s_Z^2
\right)+
2\lambda^{uL}+2\lambda^{uR}+\lambda^{dL}+\lambda^{dR},\nonumber\\
&&g_V^n=-\frac12\rho_{\nu N}^{NC}+
\lambda^{uL}+\lambda^{uR}+2\lambda^{dL}+2\lambda^{dR}\,.
\label{vcouplings}
\end{eqnarray}

Besides string inspired models, we also consider left-right symmetric
models. In this case the coupling constants in Eq. (\ref{couplings})
can be expressed as~\cite{Polak:1991pc}

\begin{eqnarray}
&&
f^{uL}=\rho_{\nu N}^{NC} A 
\left(
\frac12-\frac23 \hat\kappa_{\nu N} \hat s_Z^2
\right) - B \frac23 \hat s_Z^2 + \lambda^{uL},\nonumber\\
&&
f^{dL}=\rho_{\nu N}^{NC} A
\left(
-\frac12+\frac13 \hat\kappa_{\nu N} \hat s_Z^2
\right) + B \frac13 \hat s_Z^2 + \lambda^{dL},\nonumber\\
&&
f^{uR}=\rho_{\nu N}^{NC} A 
\left(
-\frac23 \hat\kappa_{\nu N} \hat s_Z^2
\right) + B \left(\frac12-\frac23 \hat s_Z^2
\right)  + \lambda^{uR},
\label{couplings-lr}\\
&&
f^{dR}=\rho_{\nu N}^{NC} A 
\left(
\frac13 \hat\kappa_{\nu N} \hat s_Z^2
\right) +  B \left(-\frac12+\frac13 \hat s_Z^2
\right)  +
\lambda^{dR}\nonumber
\end{eqnarray}
where
\begin{eqnarray}
&&
A = 1 + \frac{\hat s_Z^4}{1- 2  \hat s_Z^2} \gamma \nonumber\\
 &&
B = \frac{\hat s_Z^2 (1-\hat s_Z^2)}{1- 2  \hat s_Z^2} 
\label{ab}
\end{eqnarray}

\subsubsection{Neutrino-electron scattering coupling constants}

For the case of neutrino-electron scattering the total Lagrangian has the
form
\begin{equation}
\label{lagrangian2}
{\cal L}_{\nu e}^{NC}=-\frac{G_F}{\sqrt{2}}
\sum_{\alpha,\beta=e,\mu,\tau}
\left[\bar\nu_\alpha \gamma^\mu (1-\gamma^5)\nu_\beta\right]
\left(
f^{e L}\left[\bar e\gamma_\mu (1-\gamma^5) e\right] +
f^{e R}\left[\bar e\gamma_\mu (1+\gamma^5) e\right] 
\right),
\end{equation}
with $f^{e L,R}=g_{L,R}\pm \varepsilon^{L,R}$, and 
\begin{eqnarray}
\label{couplingsnewz2}
\varepsilon^{L}&=& 2 \gamma \sin^2\theta_W \rho_{\nu e}^{NC}
\left({3 c_\beta \over 2 \sqrt{6}}+{s_\beta \over 3}\sqrt{5 \over 8}
\right)^2
\nonumber \\
\varepsilon^{R}&=& 2 \gamma \sin^2\theta_W \rho_{\nu e}^{NC}
\left({c_\beta \over 2 \sqrt{6}}-{s_\beta \over 3}\sqrt{5 \over 8}
\right) \left({3 c_\beta \over \sqrt{24}}+{s_\beta \over 3}\sqrt{5 \over
8}
\right) .
\end{eqnarray}
As in the previous subsection, here $\gamma=(M_Z/M_{Z'})^2$.  With this 
Lagrangian, the neutrino-electron scattering will keep the same 
form 
\begin{equation}
\frac{d\sigma}{dT} = \frac{2 G_F m_e}{\pi}
\left[ 
g^2_L + g^2_R\left(1 - \frac{T}{E_\nu}\right)^2 - g_L g_R \frac{m_e T}{E^2_\nu}
\right]
\label{diff:cross:sec}
\end{equation}
with the only difference that now the coupling constants $g_{L,R}$ will be 
defined as 
\begin{eqnarray}
\label{gLgR}
g_L &=& \frac12 + \sin^2\theta_W + \varepsilon^{L}\\
g_R &=&    \sin^2\theta_W + \varepsilon^{R}.
\end{eqnarray}

For the left-right symmetric case, we can express the coupling constants as 
\begin{eqnarray}
\label{gLgR-lr}
g_L^{LR} &=&  Ag_L + Bg_R\\
g_R^{LR} &=&  Ag_R + Bg_L 
\end{eqnarray}
where $A$ and $B$ were defined in Eq.~(\ref{ab}).

\subsubsection{Future sensitivity}

\begin{figure}
\includegraphics[width=0.9\textwidth]{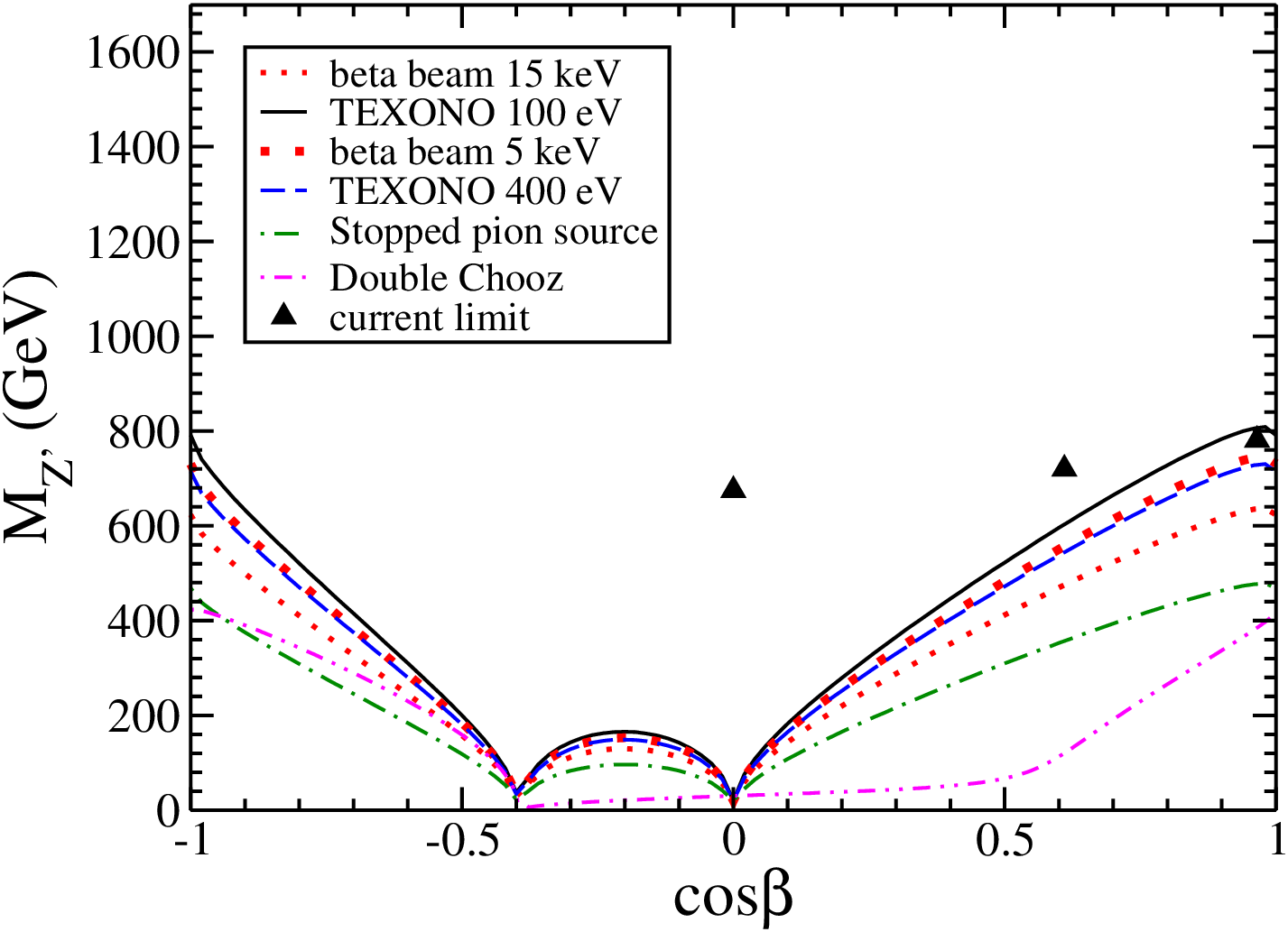}
\caption{
Sensitivity, at 95\% CL, to different extra neutral gauge boson
coming from $E_6$ models. We consider the case of the TEXONO proposal
for an energy threshold of 100~eV (solid line) and 400~eV (dashed
line); the case of a future stopped-pion source (dash-dotted line)
and a beta-beam source with energy threshold of 15 keV (bold dotted
line) and 5 keV(dotted line). Finally, the Double Chooz sensitivity is
also shown (dashed double dotted).  The current limits
(triangles) are also shown for comparison. 
}\label{fig1}
\end{figure}

In order to compute the expected $Z'$ mass limit that these experiments could
get, we consider that the future experiment will measure exactly the Standard
Model prediction, and we add the systematic error in quadratures to the
statistical one.  With these hypothesis we can compute the 95\% C.L. bound
reachable at these future experiments after one year of data taking.

We make this computation for the string inspired models for all possible
values of $\cos\beta$ considering the detector characteristics explained in
the previous section. The results are shown in Fig.~\ref{fig1}, where we also
show, for comparison, the current constraints at 95\% C.L.~\cite{Yao:2006px}.
Note that the expectations for the Double Chooz experiment are in a
qualitative agreement with similar analysis done before the MUNU experiment in
Ref.~\cite{Moretti:1998py}.
For the left-right symmetric case the expected sensitivity is shown in
Table \ref{table2}. 

\begin{table}
\begin{tabular}{|c|c|c|c|c|c|c|c|}
\hline
Experiment & Texono  & Beta beam & Beta beam & Texono  & Stopped & 
  Double Chooz & current limit \\
           & 100 eV  &   5~keV  & 15~keV    &  400 eV  & pion source & 
               &               \\
\hline\hline
Sensitivity & 450 & 419 & 358 & 406 & 251 & 565 & 860~\cite{Cheung:2001wx} \\
\hline
\end{tabular}
\caption{Expected sensitivity at 95 \% C. L., in GeV, for the mass of
a left-right symmetric model extra gauge boson. We consider five
different experimental proposals. The current limit is also shown for
comparison. }
\label{table2}
\end{table}

From Fig.~\ref{fig1} it is possible to see different phenomenological
aspects.  First, the $\chi$~model ($\cos\beta = 1$) is the most
sensitive for low energy neutrino experiments. Second, for the
coherent neutrino-nucleus scattering case, the $\psi$ model
($\cos\beta=0$) is in the opposite situation. This behavior is clear
from Eq.~(\ref{couplingsnewz}), that for this specific value the
corrections to the Standard Model Lagrangian cancel. A similar
property arises both in the case of coherent neutrino-nucleus
scattering as well as in anti-neutrino-electron scattering for
$cos\beta= -\sqrt{5/32}$.  These features of different specific models
seems to discourage the search for this type of new physics in low
energy neutrino experiments, since only a few models can give a
significant signature.  However, in the case of a positive signature
in LHC we can expect its confirmation in this kind of experiments, or
their non-observation in the case of other specific models, providing
in any case indirect complementary information.

\begin{figure}
\includegraphics[width=\textwidth]{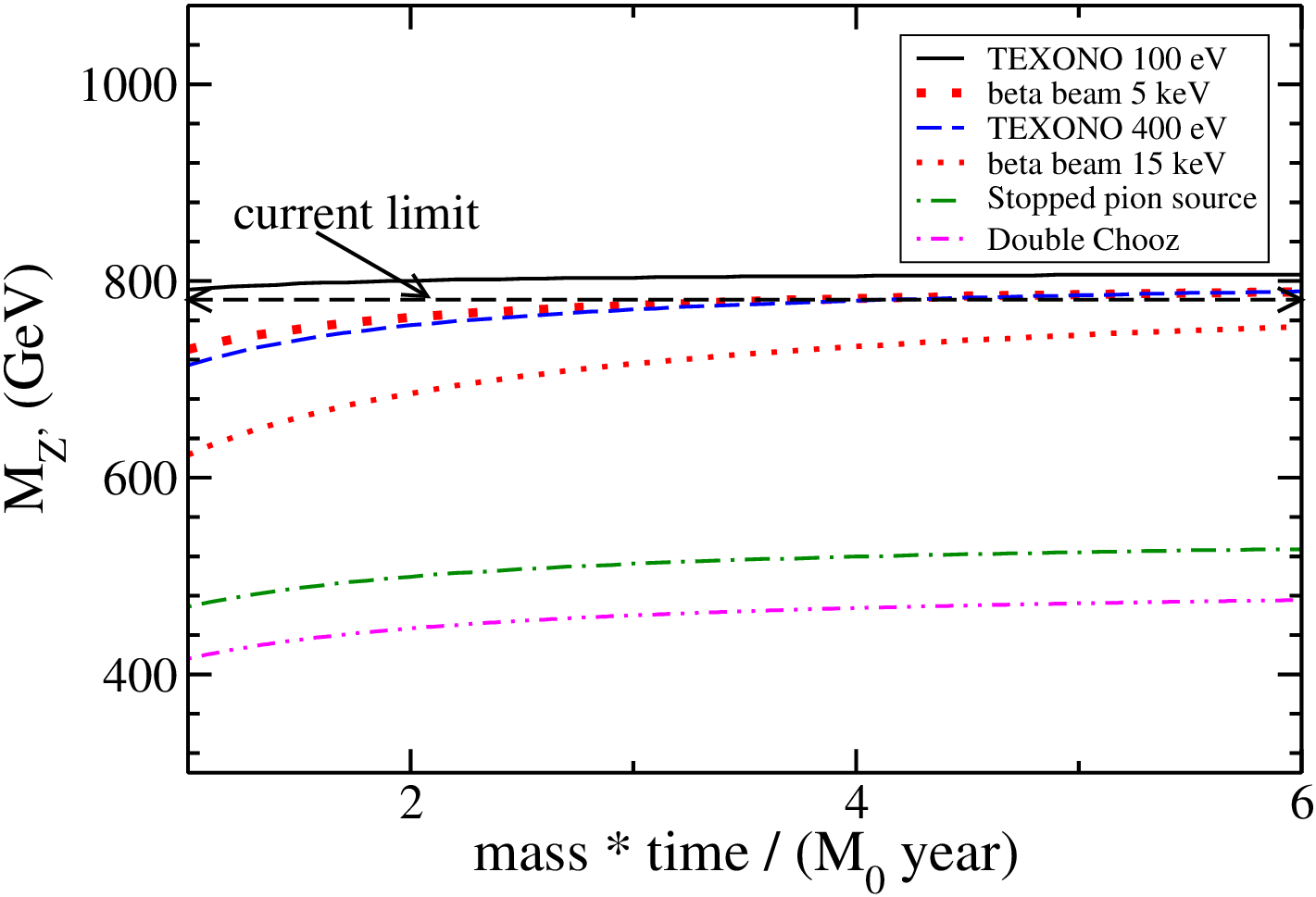}
\caption{
Sensitivity, at 95 \% C. L., to an extra $\chi$ neutral gauge boson
coming from $E_6$ models for different experimental setups. The
dependence on the size of the detector and time of running is shown.
}\label{fig2}
\end{figure}

In order to test how the sensitivity to an extra Z signal could change
with an upgraded version of these proposals, we show in
Fig.~\ref{fig2} the improved sensitivity for each proposal in the case
of an increase in mass or time exposure, which reduces the statistical
error.

We can see that in the case of extra gauge boson $Z'$ the neutrino
experimental proposals could only give a complementary information to the
current Tevatron constraints~\cite{Yao:2006px}.

\subsection{Leptoquark models}

\begin{figure}
\includegraphics[width=\textwidth]{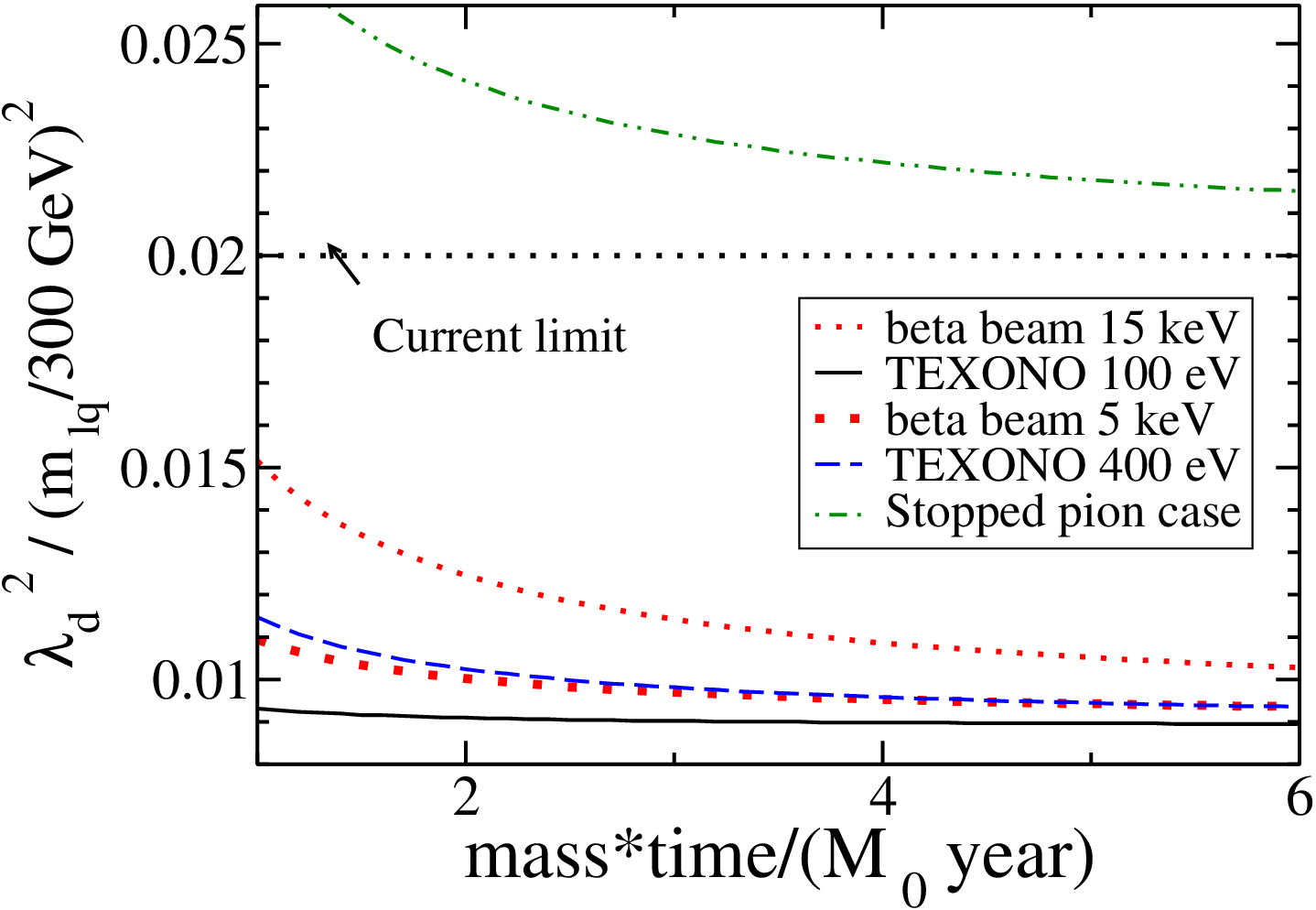}
\caption{Sensitivity, at 95 \% C. L. to a vector leptoquark coupling
for different experimental setups. The limit on the coupling
$\lambda_d$ will depend on the leptoquark mass $m_{lq}$ that here is
chosen to be 300 GeV in agreement with current literature. The
dependence on the size of the detector and time of running is also
shown.}
\label{fig4}
\end{figure}
A leptoquark is a scalar or vector boson that couples to a lepton and
a quark. There are no such interactions in the SM, but they are
expected to exist in various extensions of the
SM~\cite{Yao:2006px}, such as the Pati-Salam model~\cite{Pati:1974yy},
grand unification theories based on
$SU(5)$~\cite{Georgi:1974sy,Dorsner:2005ii} and
$SO(10)$~\cite{Fritzsch:1974nn} gauge groups and extended technicolor
models~\cite{Farhi:1980xs}. 

The leptoquark contribution effectively (in 4-fermion approximation) can
be written as~\cite{Davidson:1993qk}
$$
\varepsilon^{uV}=\frac{\lambda_u^2}{m_{lq}^2}\frac{\sqrt{2}}{4G_F}
$$
$$
\varepsilon^{dV}=\frac{\lambda_d^2}{m_{lq}^2}\frac{\sqrt{2}}{4G_F},
$$
where $\lambda_u$, $\lambda_d$ are couplings, $m_{lq}$ is leptoquark
mass. This parameterization is given for vector leptoquarks.  In the
case of scalar leptoquarks, our results should be multiplied by a
factor 1/2~\cite{Davidson:1993qk}.

\begin{table}
\begin{tabular}{|c|c|c|c|c|c|c|}
\hline
Experiment & Texono & Beta beam &  Texono & Beta beam & Stopped  & 
  current  \\

 & 100 eV& 5 keV  & 400 eV & 15 keV & pion source &  constraint\\
\hline\hline
Sensitivity & 894 & 805 & 805 & 684 & 546 & 298~\cite{Chekanov:2003af} \\
\hline
\end{tabular}
\caption{Expected 95 \% C L leptoquark mass sensitivity, in GeV, for
future low energy neutrino experiments. The leptoquark effective coupling 
has been fixed to be  $\lambda_q^2/4\pi=1/137$.}
\label{tab3}
\end{table}

In case of an observation at colliders like LHC and LEP, one can
constrain directly the leptoquark mass. The expected sensitivity for 
LHC could be as high as 1.6 TeV~\cite{Belyaev:2005ew}. However, for indirect
observations, like our low energy 4-fermion case, one can constrain
only the combination ${\lambda_q^2}/{m_{lq}^2}$. An extensive list of
constraints on the leptoquark couplings and masses is given in
Refs.~\cite{Yao:2006px,Davidson:1993qk}. The current limit for a
leptoquark which couples to the first generation of leptons and first
generation of quarks is given by 
$$
{\lambda_q^2}/{\left(m_{lq}/300\mbox{GeV}\right)^2}<0.02.
$$ 

We have calculated the sensitivity to the vector first generation
leptoquark couplings and masses which is expected at different low
energy neutrino experiments already discussed in this work. The
results are shown in Fig.~\ref{fig4} where we show the expected
sensitivity at 95 \% C. L. for each experiment and the possible
improvements if the experimental setup could run with a bigger mass
or for a longer time.

One can see that the low energy neutrino experiments are very
promising for improving the present bounds.

The sensitivities for the case of scalar leptoquark masses for
different low energy neutrino experiments are collected in Table
\ref{tab3}.  For easy comparison with the bounds given
in~\cite{Yao:2006px} we have fixed the leptoquark effective coupling
at the electroweak value: $\lambda_q^2/4\pi=1/137$ and we compute the
sensitivity of the scalar leptoquark mass at 95 \% C.L.  These
results also show a big potential for low energy neutrino experiments
to give a complementary information about leptoquark masses and
couplings.

\subsection{SUSY with broken R-parity}

\begin{figure}
\includegraphics[width=\textwidth]{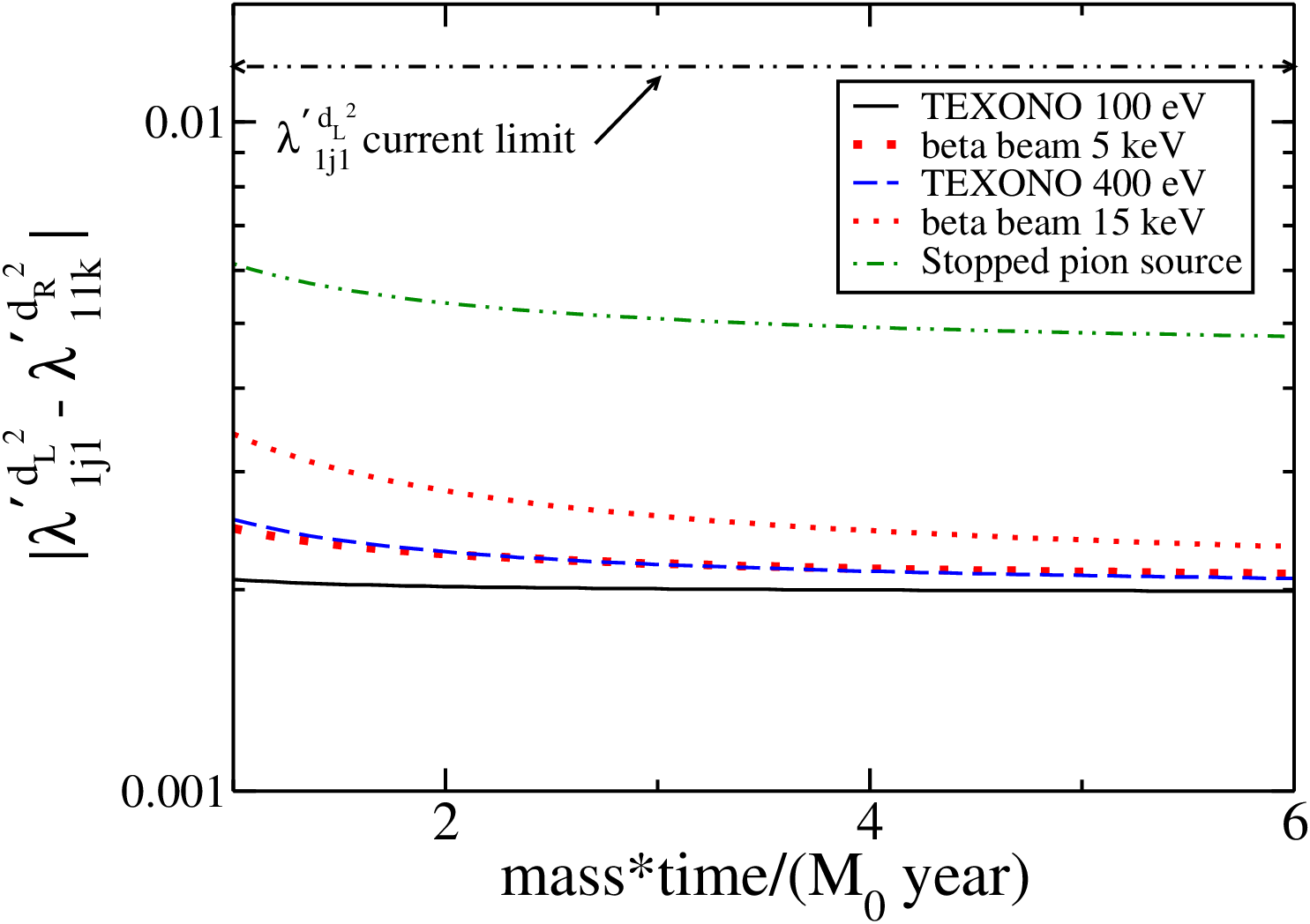}
\caption{Sensitivity, at 95 \% C. L., to neutral current $R$-parity
breaking terms for different experimental setups. The dependence on
the size of the detector and time of running is also shown as well as
the current limits. See the text for a detailed explanation of these
couplings }\label{fig5}
\end{figure}

In supersymmetric theories, gauge invariance does not imply baryon
number (B) and lepton number (L) conservation and, in general, the so
called R-parity (defined as $R=(-1)^{3B+L+2S}$ where $S$ is the spin)
is violated.
However, one has to keep the consistency with the non-observation of
fast proton decay. One may consider, for instance,
the R-parity violating MSSM (imposing baryon number conservation) with 
a superpotential that contains the following $L$- violating
terms~\cite{Barger:1989rk}:
\begin{eqnarray}
\lambda_{ijk}L_{L}^i L_{L}^j \bar E_{R}^k  \nonumber \\
\lambda_{ijk}^\prime L_{L}^i Q_{L}^j \bar D_{R}^k   ,
\end{eqnarray}
where we use the standard notation, $L_L, Q_L, \bar E_R$, and $\bar
D_R$ to denote the chiral superfields containing the left-handed
lepton and quark doublets and the right-handed charged-lepton and
$d$-quark singlets respectively; $i,j,k$ are generation indices. A
lepton-Higgs term ($LH$) can also be  included in the superpotential,
but it can be rotated away through an appropriate redefinition of the
superfields.

At low energies, the heavy Supersymmetry particles can be integrated out
and the net effect of the $R$-breaking interactions is to generate
effective 4-fermion operators involving the lepton and quark fields.

By considering the case where a single Yukawa coupling (with one
flavor structure) is much larger than the others, the effective
four-fermion operator generated by $L_L^i Q_L^j \bar D_R^k$ takes the
same form as in Eq. (\ref{lagrangian}) with the new couplings
~\cite{Barger:1989rk,Chemtob:2004xr}:
\begin{eqnarray}
&&
f^{uL}=\rho_{\nu N}^{NC}
\left(
\frac12-\frac23 \hat\kappa_{\nu N} \hat s_Z^2
\right) (1-r_{12k}(\tilde e_{kR}) ) 
+ \lambda^{uL} \nonumber\\
&&
f^{dL}=\rho_{\nu N}^{NC}
\left(
-\frac12+\frac13 \hat\kappa_{\nu N} \hat s_Z^2
\right) (1-r_{12k}(\tilde e_{kR}) ) 
+ \lambda^{dL}
-r'_{11k}(\tilde d_{kR}), 
\nonumber\\
&&
f^{uR}=\rho_{\nu N}^{NC}
\left(
-\frac23 \hat\kappa_{\nu N} \hat s_Z^2
\right) 
(1-r_{12k}(\tilde e_{kR}) ) 
+ \lambda^{uR},
\label{barger}\\
&&
f^{dR}=\rho_{\nu N}^{NC}
\left(
\frac13 \hat\kappa_{\nu N} \hat s_Z^2
\right) 
(1-r_{12k}(\tilde e_{kR}) )
+ \lambda^{dR} + 
r'_{1j1}(\tilde d_{jL}) ,\nonumber
\end{eqnarray}
where 
\begin{equation}
r_{ijk}(\tilde l)=\left( {M_W^2 \over g^2} \right)
\left( {|\lambda_{ijk}|^2 \over m_{\tilde l}^2} \right).
\end{equation}
The factors $ (1-r_{12k}(\tilde e_{kR}) )$ account for the Fermi
coupling constant redefinition $G_F=G_F^{SM}(1+r_{12k}(\tilde
e_{kR}))$ that arise from the modification to the $\mu$ decay due to
$R$-breaking interaction.  
Since the value of the Fermi constant comes from muon decay experiments, we
can not get any information on the charged current SUSY parameters and we
should concentrate only in the neutral current corrections.  As already
mentioned in a previous section, a different approach has also been
considered, that is the direct detection of $\tau$ leptons in a nearby
detector~\cite{Agarwalla:2006ht}.

From Eq.~(\ref{barger}) we can see that the $R$ breaking terms appear
both in the $f^{dL}$ and $f^{dR}$ couplings. We take into account this
correlation and we show in Fig.~\ref{fig5} the possible future
sensitivity at 95 \% C. L. of the neutrino nucleus coherent
experiments to the parameter

\begin{equation}
\lambda^{\prime d_L^2}_{1j1} - \lambda^{\prime d_R^2}_{11k} =  
  {|\lambda^\prime_{1j1}|^2 \over 
  \left( {m_{\tilde d_{jL}}^2\over 100 {\rm GeV}} \right)}
 -
  {|\lambda^\prime_{11k}|^2 \over 
  \left( {m_{\tilde d_{kR}}^2\over 100 {\rm GeV}} \right)} .
\end{equation}
As in previous sections, the possible improvements if the experimental
setup could run with a bigger mass or for a larger time is shown in
Fig.~\ref{fig5}.
The current constraints for these parameters are given by
$\lambda^{\prime d_L^2}_{1j1}\leq 0.0121$ and $\lambda^{\prime
d_R^2}_{11k}\leq 0.0001$~\cite{Chemtob:2004xr}. Stringent constraints
exist for specific values of $k$ and $j$, for instance, from
neutrinoless double beta decay~\cite{Hirsch:1995zi} in the particular
case $k=j=1$ ($\lambda^{\prime d_{L,R}^2}_{111}\leq 1.5\times
10^{-7}$). We can neglect the $\lambda^{\prime d_R^2}_{11k}$ parameter
and conclude that the perspectives to improve the sensitivity to
$\lambda^{\prime d_L^2}_{1j1}$ are quite promising for this type of
experiments.

\section{Conclusions}

We have shown that low energy neutrino experiments could provide
independent and complementary information on $Z'$, leptoquark 
masses and couplings and R-parity violating SUSY interactions.
We have calculated the potential of various future low energy neutrino
experiments to either confirm the discovery of extra heavy gauge bosons
at LHC or to constrain their masses.

As concrete coherent neutrino-nuclei interaction proposals, we have
discussed the TEXONO case, the stopped pion source with a noble gas
detector and the beta beams. In the neutrino-electron-scattering case
we have concentrated on the Double Chooz experiment. 
We have found that a coherent neutrino-nuclei scattering using reactor
neutrinos, such as the TEXONO proposal, or a beta-beam neutrino
source, could have a high sensitivity to new interactions coming from
leptoquarks or $R$-parity breaking SUSY, and we showed that the case
of a stopped-pion source experiment could also improve the current
$R$-parity breaking SUSY constraints. On the other hand, for this kind
of experiments an improved constraint to extra heavy neutral gauge
bosons seems to be difficult.

For the particular case of leptoquarks, we have found that all the
discussed low energy neutrino experiments have the potential to
improve the present bound on leptoquark masses and couplings.
In particular, the sensitivity to the vector leptoquark mass is of the
order of 800~GeV, assuming an electroweak value of the coupling,
$\lambda_q^2/4\pi= 1/137$. For the case of supersymmetry with broken
R-parity, the perspectives to improve the constraint on the
$\lambda'_{1j1}$ and the corresponding mass for the $\tilde{d_L}$ are
also very promising for all the experimental setups.

Finally, we would like to remark that low energy neutrino experiments
have great potential to provide us an indirect information about high
energy physics and therefore strongly complement accelerator
experiments.

\begin{acknowledgments}
  We kindly thank for very productive discussions to G. Centenario and
  M. Hirsch. This work has been supported by CONACyT, SNI-Mexico and 
  PAPIIT project No. IN113206. 
  TIR was supported by the Marie Curie Incoming International
  Fellowship of the European Community and he also acknowledges
  partial support from the Russian foundation for basic research
  (RFBR) and from the RAS Program ``Solar activity''.  TIR thanks
  the Physics Department of CINVESTAV for the hospitality during the visit
  when part of this work was done.
\end{acknowledgments}


\end{document}